\documentclass[12pt]{extarticle}
\usepackage[utf8]{inputenc}
\usepackage[T1]{fontenc}
\usepackage[a4paper, left = 2.5cm,right = 2.5cm,top = 2.5cm,bottom = 3cm]{geometry}
\usepackage{amsmath,amssymb}
\usepackage[linktocpage=true, colorlinks=false, linkcolor = blue, citecolor = red]{hyperref}
\usepackage{amsrefs}

\newcommand{\dd}{\partial}
\newcommand{\de}{\delta}
\newcommand{\m}{\mu}
\newcommand{\n}{\nu}
\newcommand{\ls}{\left(}
\newcommand{\rs}{\right)}

\newcommand{\la}{\lambda}
\newcommand{\ka}{\varkappa}
\newcommand{\tr}[1]{\overset{{\scriptscriptstyle 3}}{#1}{}}
\newcommand{\pos}{\tr{\Pi}_{\!\!\bot}{}}

\newcommand{\po}{{\Pi_{\!\!\bot}}}
\newcommand{\ga}{\gamma}

\newcommand{\disn}[2]{$$\displaylines{\refstepcounter{equation}%
		\label{#1}\hskip 1em minus 1em #2\hfilneg}$$}
\newcommand{\nom}{\hfil\hskip 1em minus 1em (\theequation)}
\newcommand{\str}[1]{\mathrel{\mathop{\longrightarrow}\limits_{#1}}}

\begin{document}
	
	\title{Canonical description for
		formulation of embedding gravity as a field theory in a flat spacetime}
	\author{S.A. Paston, E.N. Semenova, A.A. Sheykin\\
		{\it Saint Petersburg State University, Saint Petersburg, Russia}
	}
	\date{}
	\maketitle

	\begin{abstract}
		We consider the approach to gravity in which four-dimensional curved spacetime is represented by a surface in a flat Minkowski space of higher dimension. After a short overview of the ideas and results of such an approach we concentrate on the study of the so-called splitting gravity, a form of this description in which constant value surface of a set of scalar fields in the ambient flat space-time defines the embedded surface. We construct a form of action which is invariant w.r.t. all symmetries of this theory. We construct the canonical formalism for splitting gravity. The resulting theory turns out to be free of constraints. However, the Hamiltonian of this theory is an implicit function of canonical variables. Finally, we discuss the path integral quantization of such a theory.
		
	\end{abstract}

\maketitle

\section{Introduction}
The problem of unification of gravity and quantum description of reality arose at the end of 1920s, shortly after the appearance of the quantum mechanics.
It did not take long to describe the behavior of quantum system in an exterior gravitational field. In 1929 Fock \cite{fock_tetr} and Weyl \cite{weyl_tetr} obtained a covariant form of Dirac equation using new gravitational variables (vielbein), which allowed to correctly introduce the notion of a spinor in a gravitational field.
The quantization of the gravity itself turned out to be a much more nontrivial problem.
In the 1936 Bronstein wrote \cite{bronstein}:
\begin{quotation}
	\noindent ``In the framework of special relativity (i.e., when the spacetime is pseudo-Euclidean) one can construct quite consistent quantum theory of gravity. However, in the framework of general relativity, where the fluctuations might be arbitrarily large, the situation is drastically different $\left\langle\ldots\right\rangle$ It seems hardly possible to generalize the quantum theory of gravity onto this case without the deep reformulation of classical concepts''{}.
\end{quotation}

Since the very beginning of the search for quantum gravity physicists had been divided by the question of the meaningfulness of the problem: many prominent researchers treated (and some  still treat) the quantum-theoretical description of gravitational field as impossible or senseless \cite{penrose,sardanashvili}. However, a significant part of leading field-theorists of the past century joined the quest of construction of quantum gravity. In particular, in the above-mentioned paper Bronstein deduced the difference of signs in the Coulomb and Newton laws from the quantum positions. The necessity of inclusion of flat spacetime in the theory mentioned by him had defined the direction of subsequent search.

In the 1950s the attempts of construction of quantum gravity as a theory of massless spin-2 field in a flat spacetime had started. Gupta was among the first who attempted to quantize not only the linear theory \cite{gupta} (independently of Bronstein), but also the full theory \cite{gupta2}. Later this the development of this approach was continued by Ogievetsky \cite{ogi}, who also tried to construct quantum gravity using representations of diffeomorphism group \cite{ogi2}. While Feynman delivered his famous \emph{Lectures on Gravitation}   \cite{feyngrav}, he was possibly driven by a wish to quantize gravity as a massless spin-2 field. The significant contribution to the development of quantum gravity was made by Faddeev and Popov, who proposed the method of non-Abelian field quantization with this aim in mind \cite{faddpop}.

Unfortunately, none of these attempts can be considered as successful. One of reasons of that, in our opinion, lies in the fact that giving the special status to the Minkowski space (or at least some space of constant curvature) in the framework of GR leads to serious technical issues. It therefore makes sense to ask whether it is possible to modify the theory of gravity in such a way that it is possible to use some background spacetime for the needs of quantum field theory, without marking off any metric in the description of {our} curved spacetime. This situation is somewhat similar to the abovementioned problem of description of fermions in the gravitational field through spinor representations of Lorentz group, which is absent in the metric formulation of GR. This problem is solved in the tetrad formulation of GR, where the Minkowski metric appears as the metric of the tangent space. Another possibility to include the background metric into a theory is the equipping the original manifold with two independent metrics. The subsequent development of this idea led to the appearance of the so-called bimetric theories. However, such theories suffer from their own problems \cite{boulware}, whose solutions took almost 40 years \cite{drgt}.
The additional drawback of above-mentioned attempts is their perturbative character. Indeed, the most nontrivial effects of quantum gravity, e.g., the ones that are related to black holes, have non-perturbative nature and therefore are inaccessible by perturbative methods.

However, the idea of inclusion of flat spacetime in the theory is not discredited by failure of these attempts. In fact, the drawbacks mentioned above are associated mainly with theories in which flat ambient spacetime is treated as {our} observable spacetime, whereas other variants of its appearance are also possible. In particular, flat spacetime arises in the string framework, which allows us to construct the quantum theory of gravity (a detailed historical review of this topic can be found in \cite{rickles}). Due to this possibility the string theory can be considered as alternative to QFT \cite{marshakov}.

In 1975 Regge and Teitelboim \cite{regge} proposed a string-inspired approach to gravity in which curved spacetime is described in terms of four-dimensional surface locally isometrically embedded in a flat ambient spacetime of higher dimension. This modification of gravity is called embedding theory\cite{statja18}, Regge-Teitelboim (RT) model \cite{rojas06,estabrook2009}, geodetic brane gravity \cite{davkar} or simply embedding gravity \cite{statja51}.

In the present paper we (after a short reminder of the ideas and results of embedding gravity, see Section \ref{embgr}) will concentrate to the discussion of canonical form of the variant of embedding gravity proposed in \cite{statja25}, which is formulated as a field theory in flat ambient spacetime. The Section \ref{razb} is devoted to the formulation of this theory (``splitting gravity''). In the Section \ref{razbkan} we construct its canonical formulation. To do this, we firstly perform an ADM-like decomposition of the Lagrangian (which corresponds to the Einstein-Hilbert (EH) one written in the variables of splitting gravity) to extract generalized velocities. Then the obtained expression is used for construction of canonical formulation w.r.t. time-like coordinate of flat ambient spacetime. In Section  \ref{funk} we discuss the possibility of path integral quantization of splitting gravity in canonical variables.	
\section{The ideas and results of embedding gravity}\label{embgr} In the framework of embedding gravity the spacetime is assumed to be not just a pseudo-Riemannian space with a metric $g_{\m\n}(x^\ga)$
($\m,\n,\ldots=0,1,2,3$), but a surface $\mathcal M$ in an $N$-dimensional Minkowski spacetime (as in the string theory which was the main source of inspiration of Regge and Teitelboim \cite{regge}). All geometric characteristics of this surface, including metric, can be expressed through embedding function  $y^a(x^\mu)$ ($a,b,\ldots=0,\ldots,N-1$), which defines $\mathcal M$ parametrically. The metric is assumed to be induced by the flat metric of the ambient spacetime $\eta_{ab}$, i.e., defined in the following way:
\begin{align}
g_{\mu\nu}=\eta_{ab} (\partial_\mu y^a) (\partial_\nu y^b). \label{metric}
\end{align}

The case $N=10$ is usually considered since in this case the number of new variables $y^a(x^\mu)$ coincides with the number of independent components of the old variable $g_{\mu\nu}$. However, for the greater generality we will not fix the dimension $N$ of the ambient space in this paper.

In the variational principle for embedding gravity the embedding function has to be varied instead of metric. If the action is chosen to be the Einstein-Hilbert action with some material contribution ${\cal L}_m$ (we use the signature $+---$)
\begin{align}\label{deistv}
S=\int d^{4}x \sqrt{-g}\ls-\frac{1}{2\ka}\, R+{\cal L}_m\rs,
\end{align}
then, after the substitution of \eqref{metric} in the action and variation w.r.t.~$y^a$, the Regge-Teitelboim equations appear:
\begin{align}\label{urRT}
\partial_\mu \ls\sqrt{-g}(G^{\mu\nu}-\varkappa\, T^{\mu\nu})\partial_\nu y^a\rs=0,
\end{align}
where $G^{\mu\nu}$ is the Einstein tensor and  $T^{\mu\nu}$ is the energy-momentum tensor.
These equations are obviously more general than the Einstein ones, though all solutions of Einstein equations also solve RT equations.

Since the purpose of the initial development of RT approach was not to extend the Einsteinian dynamics, but rather to reformulate it in new variables, the dynamics was often (since the original work \cite{regge}) restricted by so-called Einsteinian constraints:
\begin{align}
G^{0\mu}-\ka\, T^{0\mu}=0.\label{ein_constr}
\end{align}

The correctness of the introduction of the additional constraints was discussed, among other aspects of the theory, in the paper \cite{deser}. Later it was shown \cite{statja18} that if 
the initial values are restricted by Einsteinian constraints, the entire dynamic becomes Einsteinian in all moments of time. The analysis of RT equations in the assumption of homogeneity and isotropy of the universe shows that if the embedding function has analogous symmetry and initial values are not fine-tuned the dynamics of the universe in the post-inflational period becomes almost Einsteinian
\cite{statja26}. It should be noted, though, that another choice of initial values corresponds to the existence of the necessary amount of dark matter  \cite{davids01}. Note also that in some cases (e.g., in the Schwarzchild-like one) extra solutions vanish after the imposition of boundary  conditions \cite{statja33} instead of initial ones \eqref{ein_constr}.

Regge and Teitelboim proposed the embedding gravity in the hope that the natural appearance of the flat ambient spacetime in the theory could help to resolve some problems of the canonical quantization of gravity. After the papers \cite{regge,deser} the procedure of isometrical embedding in application to gravity (including quantization) discussed by several authors, see, e.g., \cite{pavsic85let,maia89,estabrook1999,davkar,estabrook2009,faddeev},
and also \cite{tapiaob} for a list of additional references.
To move towards the quantization it is important to construct a canonical formulation of the theory, both with Einsteinian constraints \eqref{ein_constr} and without them. Besides the original paper \cite{regge}, there are some works devoted to this topic, e.g., \cite{tapia,frtap,statja18,rojas09,statja24,statja35,statja44,rojas20}.

The extension of dynamics appearing in the embedding gravity can be used in classical gravity to explain the effects of dark energy and dark matter. In the assumption of homogeneity and isotropy of the universe this idea was studied in \cite{davids97,davids01}. One can drop this assumption if one reformulates embedding gravity as GR with some additional matter \cite{statja48,statja51}, analogously to the reformulation of the mimetic gravity \cite{mukhanov} in the form of GR plus mimetic matter \cite{Golovnev201439}. This similarity between embedding gravity and mimetic gravity is due to that fact that in both theories the modification of theory occurs as a result of substitution of the expression for the metric, which contains derivatives of some new variables, into EH action. After such modification of the theory, which is a particular case of differential transformations of field variables, the extension of dynamics occurs in most cases \cite{statja60}.

It should be noted that in the framework of embedding theory the description of gravity is quite different from other interactions, which are described by fields---functions of a point in a flat spacetime. The embedding function $y^a(x^\m)$, which is the independent variable in the embedding theory, in turn, is not   a function of the point in a flat spacetime. Because of this fact the proceture of canonical quantization is still difficult because after the quantization $y^a(x^\m)$ becomes an operator-valued function, as well as metric (according to \eqref{metric}), so the situation is similar to the quantization of GR in its usual formulation.
To solve this problem one might try to reformulate the theory in terms of another variables which are, in some sense, dual to the original variables of the embedding theory.

Since in embedding theory a physical sense is given to the shape of the surface (corresponding to the equivalence class of the embedding functions $y^a(x^\m)$ related to each other by diffeomorphisms), one can obtain an alternative way to define it without the introduction of coordinates on the surface. Such way, which is based on the possibility to define a surface algebraically rather than parametrically, was proposed in \cite{statja25}. The splitting gravity appearing on this way turns out to be the field theory in a flat ambient space of higher dimension. It should be noted that the diffeomorhic invariance in this theory is absent as it is already formulated in the (in some sense) diffeomorhic-invariant quantities.
\section{The splitting gravity: Regge-Teitelboim approach in the form of a field theory}\label{razb}The idea, on which splitting gravity is based, is the following. A four-dimensional surface $\mathcal M$, which is the main object of study in the embedding gravity, can be defined in a coordinate-free fashion. To do that, one must define  a set of $N-4$ scalar fields $z^A$, $A=0,\ldots,N-5$, in $N$-dimensional Minkowski spacetime. The level sets of these fields are given by equations
\begin{align}\label{s0}
z^A(y^a)=const,
\end{align}
which define a four-dimensional surface in $N$-dimensional flat spacetime. All geometric characteristics of these surfaces, which are not connected to the choice of coordinates on them, can be written in terms of  $z^A(y^a)$. The term ``splitting gravity'' is chosen due to the fact that conditions \eqref{s0}, in fact, ``split'' the whole $N$-dimensional spacetime into a family of four-dimensional surfaces corresponding to the different values of constants. {Note that there is another approach to gravity which is called ``gravitational splitting''} \cite{gravsplit}. It deals with the factorized (``splitted'' in the sense of splitting of operator algebras) Hilbert space of a system with gravitational interaction. The splitted space in our paper is not a space state, but rather the bulk space, so our approach is completely different.

The description of surfaces in terms of $z^A(y^a)$ is, in some sense, dual to their description in terms of embedding function $y^a(x^\m)$, but without the necessity of the introduction of coordinates on the surfaces. If only the shape of the surfaces is assumed to be observable, then some new symmetry appears instead of diffeomophic invariance. Indeed, the way in which the function $z^A(y^a)$ ``splits'' the flat spacetime into the system of surfaces is not altered by the substitution
\begin{align}\label{s1}
z^A(y^a)\rightarrow z'^A(y^a)=f^A\ls z^B(y^a)\rs.
\end{align}

In splitting gravity this substitution, which ``renumerates''{} the surfaces, plays the role which is analogous to the one that diffeomorphisms plays in the GR. The difference between them lies in the fact that diffeomorphisms are acting in the space of \emph{arguments}  
of fields, whereas transformations \eqref{s1} are acting in the space of \emph{values} of fields.

The question about the nature and properties of this ``renumeration symmetry'' \eqref{s1} is nontrivial.  On one hand, it is not a global symmetry in the usual sense, since it is defined not by a finite number of parameters, but rather by a set of functions $f^{A}$. On the other hand, it is not a local one either, since, firstly, the renumeration transformation does not change the properties of the surfaces locally (i.e., independently in an each point), but rather thansforms the set of surfaces as a whole. Secondly, the number of variables, on which the transformation variables $f^A$ depend, is nevertheless smaller than the number of parameters, on which the field $z^A$ itself depends. Since this symmetry is related to the arbitrariness of definition of surface through quantities $z^A$, it should be considered as gauge one, though it does not lead to the appearance of any constraints, see below.

By analogy with tensors w.r.t. usual diffeomorphisms, one can introduce ``renumeration tensors''---the quantities which have upper and lower indices and transforms according to ``tensor''~law
\disn{sn1}{
	q'^A=\frac{\dd z'^A}{\dd z^B}q^B,\qquad
	q'_A=\frac{\dd z^B}{\dd z'^A}q_B,\qquad
	\ldots
	\nom}
under the transformations \eqref{s1}.
It is easy to see that the quantity
\disn{s2}{
	v_a^A=\dd_a z^A
	\nom}
is a renumeration tensor w.r.t. index $A$.

After differentiation of the renumeration tensors, as usual, one can obtain quantities which are not tensors. However, straightforward calculation shows that tangent derivative
\disn{s3}{
	\bar\dd_a\equiv \Pi^b_a\dd_b
	\nom}
(here $\Pi^b_a$ is an orthogonal projector on the tangent space at the given point, see below for the exact formula)
applied to a tensor gives a tensor again. Therefore in splitting theory the tangent derivative $\bar\dd_a$ is analogous to the covariant derivative in GR and embedding theory.

By analogy with the construction of the non-square vielbein by the differentiation of the embedding function $e^a_\m=\dd_\m y^a$ (see details in \cite{statja18}), of which the metric \eqref{metric} consists, in the splitting theory the derivative of $z^A$, i.e., $v_a^A$, is also some non-square vielbein. Using it, one can construct the following quantity:
\begin{align}\label{s5-3}
w^{AB}= v_a^A v_b^B \eta^{ab},
\end{align}
which can be used as a ``metric'' for the renumeration tensors defined above. In particular, it (together with its inverse $w_{AB}$) allows to raise and lower the indices like $A,B,\ldots$ It must though be ke in mind that $w^{AB}$ is not a metric of any submanifold since $w^{AB}$ depends on $y^a$ and not just on $z^A$.

As in the embedding theory, where non-square vielbein allows to construct projectors on the tangent ($\Pi^b_a$) and  orthogonal ($\po^b_a$) spaces to $\mathcal M$ at a given point, in the splitting theory one can also construct these projectors of  $v_a^A$:
\begin{align}\label{s4}
\po_{ab}=  v_a^A v_b^B w_{AB}, \qquad \Pi^a_b=\de^a_b-\po^a_b,
\end{align}
see details in \cite{statja25}.

The only renumeration tensor
which is linear w.r.t. second derivatives of $z^A$ is the following:
\begin{align}\label{s5}
\hat{b}^A{}_{bc}\equiv -\bar\dd_b v^A_c=v^A_e \partial_c \Pi^e_{b}.
\end{align}

After ``transferring''{} into the ambient space w.r.t. index $A$, the resulting quantity can be expressed through projectors \cite{statja42}:
\begin{align}\label{s6}
\hat{b}^d{}_{bc}=v^d_A\hat{b}^A{}_{bc}=\po^d_e \partial_c \Pi^e_{b}=\Pi^e_b \partial_c \Pi^d_{e}
\end{align}
(see the properties of projectors used here in \cite{statja25}).
The projection of this quantity is the well-known geometric characteristic of the surface $\mathcal M$ called the second fundamental form
{which can be ``transferred''{} into the ambient space}:
\begin{align}
b^d{}_{ba}=\Pi^c_a \hat{b}^d{}_{bc}=\Pi^e_b\Pi^c_a \partial_c \Pi^d_{e}.
\end{align}

The Riemann curvature tensor (together with Ricci tensor and Ricci scalar) can be constructed of the second fundamental form:
\vspace{6pt}
\disn{s2aa}{
	R_{abcd}=b^e{}_{ac} b_{ebd} -b^e{}_{ad} b_{ebc}.
	\nom}

It is easy to see that one cannot construct a nontrivial scalar w.r.t.~\eqref{s1} and ambient Lorentz group using only the first derivatives of $z^A$. Therefore the action must also consist of second derivatives and thus of the quantity  \eqref{s5}. Since it has three indices, the simplest nontrivial scalars are contractions of two such quantities \cite{statja42}. Among these scalars one can pick out the
Ricci scalar:
\begin{align}
\hat b^e{}_{a}{}^a \hat b_{eb}{}^b -\hat b^e{}_{a}{}^b \hat b_{eb}{}^a=
b^e{}_{a}{}^a b_{eb}{}^b -b^e{}_{a}{}^b b_{eb}{}^a=R.
\end{align}

However, the Ricci scalar alone is not enough to obtain an analogue of the EH action. In the assumption of the absence of interaction between surfaces one can write down the action of the theory as an integral of the each surface action \eqref{deistv} over $z$:
\begin{align}\label{ss5}
S=\int dz\, d^{4}x \sqrt{-g}\ls-\frac{1}{2\ka}\, R+{\cal L}_m\rs.
\end{align}

Coordinates $x^{\mu}$ used here are introduced temporarily and will disappear in the final answer. To see that, one should notice that the transition from the integration w.r.t. a set $\{x^\mu, z^A\}$, which plays the role of curvilinear coordinates in the ambient Minkowski spacetime, to the integration w.r.t. Lorentz coordinates $y^a$ in it  leads to the appearance of a Jacobian, so the action takes the form
\begin{align}\label{act}		
S=\int dy \sqrt{|w|} \ls-\frac{1}{2\ka}\, R+{\cal L}_m\rs,
\end{align}
where $w=\text{det}\,w^{AB}$, and derivatives w.r.t.~$x^\m$, which are present in  $\mathcal{L}_m$, have to be substituted by tangent derivatives  \eqref{s3} (see details in \cite{statja25}).
If the action is chosen as such, it turns out that the fluctuations of matter propagate only along the surfaces $\mathcal M$ despite the fact that the matter fields are initially defined in a whole ambient spacetime.

The uniqueness of the weight multiplier $\sqrt{|w|}$ in the action of splitting gravity can be noticed without the introduction of auxiliary coordinates  $\{x^\mu, z^A\}$, if one considers the brane theory in this~approach:
\begin{align}\label{w}
S=\int dy\, F(w),
\end{align}
where $F(w)$ is an arbitrary function. Writing down the the corresponding  equation of motion and requiring it to be the same as the known equation of motion of a brane 
\begin{align}\label{sqrt}
b^a{}_{c}{}^{c}=0,
\end{align}
one can find \cite{statja42} that the square root is the only type of function  $F(w)$ which leads to \eqref{sqrt}.

The variation of \eqref{act} w.r.t. $z^A$ gives the following equations:
\begin{align}\label{ss2}
(G^{ab} - \varkappa\, T^{ab}) b^A{}_{ab} = 0,
\end{align}
which reproduce one of the forms of RT equations (\ref{urRT}) which govern the dynamics of the each surface $\mathcal M$.
The Einstein tensor $G_{ab}$ and matter EMT $T_{ab}$ used here is a result of ``transferring''{} of the corresponding quantities used in (\ref{urRT}) into the ambient space.

It is interesting to note that the equation of motion \eqref{ss2} turns out to be covariant w.r.t.
renumeration symmetry \eqref{s1} (the left-hand side of this equations is transformed according to the tensor law \eqref{sn1}), despite the fact the action
\eqref{act} is not invariant w.r.t. it, since  $w$
transforms as follows:
\disn{ss3}{
	w'=\ls\det\frac{\dd z'^A}{\dd z^B}\rs^2 w.
	\nom}

The lack of action invariance w.r.t. this symmetry in the specific case that we consider (i.e., action that is a sum of contributions of all surfaces, and each contribution does not contain the field derivatives in the directions which are normal to the surface) does not lead to the alteration of the equations of motion, because the only thing that changes under the renumeration symmetry is the weight $\Phi(z)$ that defines a magnitude with which a surface contribute to the whole action. Since our choice of the action excludes any interaction between different surfaces, the physics on the each one is unaffected by renumeration symmetry.

However, an explicit invariance of the action w.r.t. symmetries of the theory can play some role in its quantization  (e.g., in the path integral quantization). It is thus interesting to construct an alternative form of the action which is invariant w.r.t. \eqref{s1} but nevertheless corresponds to the equation of motion~\eqref{ss2}.

It turns out (as was
{proposed}
by Grad) that it can be done by introducing some auxiliary variables. The resulting action takes the form
\disn{ss4}{
	S=\int dy  \ls-\frac{1}{2\ka}\, R+\bar\dd_a\ls\Pi^a_b\xi^b\rs+{\cal L}_m\rs\la,
	\nom}
where the auxiliary variables are the scalar field $\la$, which plays the role of a Lagrange multiplier,
and some vector field $\xi^b$.
It
{can be checked}
that the variation of this action  w.r.t. all independent variables indeed leads to equations \eqref{ss2}.

To do that in a simple way, one can temporarily introduce coordinates $x^\mu$ on the surfaces, so the whole spacetime is parametrized by a set of curvilinear coordinates $\tilde{y}^a = \{x^\mu(y^a), z^A(y^a)\}$.
Then one {can} rewrite the action {\eqref{ss4}} in a form analogous to \eqref{ss5}:
\disn{ss6}{
	S=\int dz \int d^4x \sqrt{-g}  \ls-\frac{1}{2\ka}\, R+D_\mu \xi^\mu(y(x,z)) +{\cal L}_m\rs \frac{\la(y(x,z))}{\sqrt{|w|}},
	\nom}
where $\xi^\mu$ is the projection of the vector $\xi^b$ on the plane tangent to $\mathcal M$ at this point, ``transferred''{} into the Riemannian space (see details in \cite{statja18}), so $\Pi^a_b\xi^b=\xi^\mu\dd_\mu y^a$.
If the formula for a covariant derivative in the embedding framework is used (see \cite{ 	statja18}), one obtains the equation
$\bar\dd_a\ls\Pi^a_b\xi^b\rs=D_\mu \xi^\mu$, which was used in the construction of \eqref{ss6}.

The variation of the action \eqref{ss6} w.r.t. $\la$ and $\xi$ gives
\begin{align}\label{ss7}
-\frac{1}{2\ka}\, R+D_\mu \xi^\mu +{\cal L}_m =0,\\
\dd_\mu \frac{\lambda}{\sqrt{w}}=0 \quad\Rightarrow\quad \la = \Phi(z) \sqrt{w}
\end{align}
respectively, where $\Phi(z)$ is an arbitrary function. Due to the fact that the original action
{\eqref{ss4}}
does not depend on
{choice of coordinates on the surfaces $x^\mu(y^a)$},
its variations  w.r.t.  $\tilde{y}^a(y^a)$ and $z^A(y^a)$ are equivalent. Then, instead of the variation of
{action} w.r.t. {function} $\tilde{y}^a (y)$, let us vary it  w.r.t. its inverse
{function}
$y^a (\tilde{y})$, which is an equivalent operation. Using the equations \eqref{ss7}, one can obtain that, {when} the action is varied w.r.t.
{$y^a (x,z)$,} \eqref{ss6} can be replaced by
\disn{ss6a}{
	S=\int dz\, \Phi(z) \int d^4x \sqrt{-g}  \ls-\frac{1}{2\ka}\, R +{\cal L}_m\rs + \int dz\, \Phi(z) \int d^4x \sqrt{-g}\,D_\mu \xi^\mu.
	\nom}

The variation of the first term  gives the usual RT equations, whereas the second term  is a surface integral which does not contribute to the equations of motion.

Finally, it it interesting to note that the same procedure allows one to construct the field-theoretic form of not only the EH action, but also any kind of an action for some surface. The simplest example is a brane (and, in particular, a string) action, for which $R$ in \eqref{ss4} needs to be replaced by an arbitrary constant and ${\mathcal L}_m$ needs to be excluded. It is worth mentioning that another modification of the original embedding approach also provides a way to alternative description of strings and branes, see \cite{statja60}.

\section{The canonical formulation of the splitting gravity}\label{razbkan}
To construct the canonical formulation of the theory described above it is, first of all, necessary to isolate the time $y^0$ among the coordinates of the ambient spacetime $y^a$, writing the arguments of the field $z$ as $z^A(y^0,y^I)$, $I=1\ldots 9$. Therefore at a given point of time $y^0$ this field splits the $(N-1)$-dimensional space $y^0=const$ into three-dimensional surfaces $\tr{\mathcal M}$. These surfaces can be characterized by quantities  similar to the ones which were introduced in the Section \ref{razb} for four-dimensional surfaces in the $N$-dimensional spacetime:
\disn{ra11}{
	\tr v_I^A=\dd_I z^A,\qquad
	\tr w^{AB}=\tr v^A_I \tr v^B_K\eta^{IK},
	\nom}
and also $\tr{w}_{AB}$, $\tr{v}^I_A$, $\tr{\Pi}^I_K$, $\pos^I_K$, $\tr{b}^A{}_{KL}$, $\tr{R}_{IKLM}$, and so on.

To simplify the construction of the canonical formulation let us restrict ourselves to the vacuum case ${\cal L}_m=0$, as it was usually done in the construction of canonical formulation for various forms of embedding gravity (see the papers mentioned in the Introduction).
We rewrite the gravitational part of the action (\ref{act}) (omitting boundary terms) by temporary introduction of coordinates $x^\m$ on the surfaces, in the first order form.
To do this, let us recall the formula used in ADM approach which connects the four-dimensional curvature $R$ with the three-dimensional one $\tr R$:
\disn{ra11a}{
	R=\tr R+(K^i_i)^2-K_{ik}K^{ik}+\dd_i\xi^i,
	\nom}
where $i,k,\ldots=1,2,3$ and $K_{ik}$ is the second quadratic form of surface $x^0=const$ in 4D space-time.
Taking into account (see~\cite{statja18})
\begin{align}\label{K}
K_{ik}=n_a \tr{b}^a_{ik},
\end{align}
where
\disn{1.6.3}{
	\tr b^a_{ik}=\pos^a_b\dd_i\dd_ky^b,
	\nom}
and $n_a$ is a unit  vector normal to $\tr{\mathcal M}$ and tangent to ${\mathcal M}$, we obtain the following result:
\disn{ra12}{
	S=-\frac{1}{2\ka}\int dy\, \sqrt{|w|}\ls
	n_a\, n_b\;
	\tr{b}^a_{ik}\,\tr{b}^b_{lm}\ls\tr{g}^{ik}\tr{g}^{lm}-\tr{g}^{il}\tr{g}^{km}\rs+\tr{R}\rs.
	\nom}

Since all surfaces $\tr{\mathcal M}$ by definition are situated  in $(N-1)$-dimensional subspaces $y^0=const$,
we conclude that $\tr{b}^0_{ik}=0$.
The other components of $\tr{b}^L_{ik}$ can be ``transferred''{} into the
$(N-1)$-dimensional space using the formula $\tr{b}^L_{IK}=\tr{b}^L_{ik}\tr e^i_I \tr e^k_K$,
which allows to rewrite the action in a coordinate-free form:
\disn{ra13}{
	S=-\frac{1}{2\ka}\int dy\, \sqrt{|w|}\ls
	n_I\, n_K\;
	\ls\tr{b}^I{}_L{}^L\,\tr{b}^K{}_M{}^M-\tr{b}^I{}_M{}^L\,\tr{b}^K{}_L{}^M\rs
	+\tr{R}\rs.
	\nom}

Then it is necessary to determine its dependence of the velocities $\dot z^A\equiv\dfrac{\dd z^A}{\dd y^0}$.
The vector $n^a$ is a unit normal to $\tr{\mathcal M}$ which is tangent to $\mathcal M$. From that it follows, firstly, that its $(N-1)$ components satisfy the relation
\disn{ra14}{
	n_I=\tr{v}_I^A n_A,
	\nom}
(as $\tr{\mathcal M}$ are situated in $(N-1)$-dimensional subspaces $y^0=const$),
and the remaining component $n_0$ can be found using the condition of normalization:
\disn{ra15}{
	n^a n_a=1\qquad\Rightarrow\qquad
	n^0=-\sqrt{1-n_A\tr w^{AB}n_B},
	\nom}
where the minus sign is chosen for the convenience.
Secondly,  $n^a$ satisfy the relation
\disn{ra16}{
	n^a v^A_a=0.
	\nom}

Noticing that $\tr{v}_I^A=v_I^A=\dd_I z^A$, we obtain that
\disn{ra17}{
	n_A=-n^0 \tr w_{AB} \dot z^B\qquad\Rightarrow\qquad
	n^0=-\frac{1}{\sqrt{1+\dot z^A\tr w_{AB}\dot z^B}},
	\nom}
where we have used (\ref{ra14}), (\ref{ra11}) and then (\ref{ra15}).
As a result we obtain the expression for $n^I$ in terms of velocities $\dot z^A$:
\disn{ra17.1}{
	n^I=\frac{\tr v^I_A\dot z^A}{\sqrt{1+\dot z^A\tr w_{AB}\dot z^B}}.
	\nom}

The same procedure must be applied to the determinant  $w$ which is present in the action (\ref{ra13}).
Using (\ref{s5-3}), we have
\disn{ra18}{
	w^{AB}=v^A_a v^{Ba}=v^A_0 v^{B0}+\tr{v}^A_I \tr{v}^{BI}=\dot z^A\dot z^B+\tr{w}^{AB}.
	\nom}

From this it follows that
\disn{ra19}{
	w=\tr{w}\ls 1+\dot z^A\tr w_{AB}\dot z^B\rs.
	\nom}

Substituting (\ref{ra17.1}) and (\ref{ra19}) into the action (\ref{ra13}), we can write it in the following form:
\disn{ra20}{
	S=\int dy\, {\cal L}(z^A,\dot z^A),
	\qquad
	{\cal L}=\frac{1}{2} \ls
	\frac{\dot z^C \tr{w}_{CA} B^{AB}\tr{w}_{BD}\dot z^D}{\sqrt{1+\dot z^A\tr w_{AB}\dot z^B}}+
	\sqrt{1+\dot z^C\tr w_{CD}\dot z^D}\, \tr{w}_{AB}B^{AB}\rs,
	\nom}
where
\disn{ra21}{
	B^{AB}=-\frac{1}{\ka}\sqrt{|\tr{w}|}
	\ls\tr{b}^A{}_L{}^L\,\tr{b}^B{}_M{}^M-\tr{b}^A{}_M{}^L\,\tr{b}^B{}_L{}^M\rs.
	\nom}

In the form (\ref{ra20}) an explicit dependence on the velocities $\dot z^A$ is present, and the quantities $\tr w^{AB}$ and $B^{AB}$ depend only on the values of  $z^A$ at a given moment of time.

Having the splitting gravity action in the form (\ref{ra20}), it is easy to construct the generalized momenta
\disn{ra22}{
	\pi_A=\frac{\de S}{\de \dot z^A}=
	\tr{w}_{AB}B^{BE} n_E -\frac{1}{2} n_A \ls n_D B^{DE} n_E - \tr{w}_{DE} B^{DE}\rs,
	\nom}
where
\disn{ra22.1}{
	n_A=\frac{\tr{w}_{AB}\dot z^B}{\sqrt{1+\dot z^A\tr w_{AB}\dot z^B}}.
	\nom}

Note that the relation (\ref{ra22}) does not lead to any restrictions
on the coordinates $z^A$ and momenta $\pi_A$, so the canonical formulation for splitting gravity is constraint-free.
However, we have not succeeded in the obtaining of explicit expression of velocities $\dot z^A$ through $z^A$ and $\pi_A$ from (\ref{ra22}). Taking (\ref{ra22.1}), (\ref{ra15}) and (\ref{ra17}) into account, it is easy to obtain that
\disn{ra23}{
	\dot z^A=\frac{\tr{w}^{AB} n_B}{\sqrt{1-n_A\tr w^{AB}n_B}},
	\nom}
but an explicit expression for $n_A$ through $z^A$ and $\pi_A$ can be obtained only by solving the multidimensional cubic Equation (\ref{ra22}) w.r.t. it, which is not possible in the general case. 
Note that in the presence of matter the equation (even the simplest kind of it, see the example of a scalar field in \cite{statja25}) \eqref{ra22} becomes more complicated. Its general properties in this case are the same: one cannot solve it w.r.t. velocities; however, it depends on the velocities not only through $n^a$, which leads to significant complications in the subsequent calculations. The situation can be simplified by imposing the Einsteinian constraints \eqref{ein_constr},
since after that the bracketed expression in \eqref{ra22} vanishes.
Such approach for the embedding theory was studied in \cite{regge,statja18,statja24,statja35},
but its application to splitting gravity lies beyond the scope of this paper.

Denoting a solution of (\ref{ra22}) as $n_A(\pi,z)$, one can obtain a Hamiltonian {density for} the splitting gravity in terms of this implicitly defined function:
\disn{ra23.1}{
	{\cal H}=\pi_A\dot z^A-{\cal L}=
	\frac{1}{2}\sqrt{1-n_A(\pi,z)\tr w^{AB}n_B(\pi,z)}\ls n_D(\pi,z) B^{DE} n_E(\pi,z) - \tr{w}_{DE} B^{DE}\rs.
	\nom}

Therefore, despite the absence of constraints, the canonical formulation constructed here is not quite simple, because we cannot write down the explicit form of the Hamiltonian in terms of canonical variables $z^A$ and $\pi_A$. However, even an implicitly defined Hamiltonian can be used in the construction of the path integral w.r.t. canonical variables.

\section{The path integral with respect to canonical variables}\label{funk}
As a next step to the quantization of the obtained theory, let us consider the path integral for it w.r.t. canonical variables
\begin{align}\label{int}
I=\langle z''^A(y^a)|e^{-itH}|z'^A(y^a)\rangle=\int Dz D\pi\, \text{exp}\left(i\int dy\,
\ls\pi_A\dot{z}^A - \mathcal{H}(z,\pi)\rs  \right).
\end{align}

The analysis of this path integral is significantly complicated by the fact that the explicit dependence of the Hamiltonian {density} $\mathcal{H}(z,\pi)$ on canonical variables is unknown.
One can try to avoid this problem following the procedure proposed in \cite{cahill} (see also \cite{cahill2}).

Let us describe the idea of this procedure using the splitting gravity as an example. First recall that one obtains the Hamiltonian
from the Lagrangian
{by}
the Legendre transformation. Let us consider
the function $\mathcal L(z^A, \psi^A)$ instead of Lagrangian {density} $\mathcal L(z^A, \dot{z}^A)$, where $\psi^A$ is an auxiliary field, and perform the Legendre transformation w.r.t. this field. As usual, we introduce
\begin{align}\label{pii}
\pi_A(\psi,z)=\frac{\partial \mathcal L(z, \psi)}{\partial \psi^A},
\end{align}
then obtain (formally) the expression for $\psi^A(\pi,z)$ from {\eqref{pii}} and write down the Hamiltonian {density}:
\begin{align}\label{pii2}
\mathcal{H}(z,\pi)=\pi_A \psi^A(\pi,z)-\mathcal{L}(z, \psi(\pi,z)).
\end{align}

We can substitute this form  of Hamiltonian in the path integral (\ref{int}) and then make the change of variables in it, transforming the integration variable from $\pi_A$ to the auxiliary field $\psi^A$:
\begin{align}\label{int1}
I=\int Dz D\psi \left| \dfrac{\de\pi_A}{\de\psi^B} \right| \text{exp}\left(i\int dy
(\pi_A(\psi,z)\dot{z}^A - \pi_A(\psi,z) \psi^A + \mathcal{L}(z,\psi))  \right).
\end{align}

Using \eqref{pii} in this expression, we obtain
\begin{align}\label{int2}
I=\int Dz D\psi \left| \dfrac{\de^2 \mathcal L}{\de \psi^A \de \psi^B} \right| \text{exp}
\left(i\int dy\ls\ls\dot{z}^A-\psi^A\rs\dfrac{\partial \mathcal L(z,\psi)}{\partial \psi^A} + \mathcal{L}(z,\psi)\rs
\right).	
\end{align}

In case of splitting gravity it is more convenient to modify the procedure described above, using the integration not w.r.t. the generalized velocities $\dot{z}^A$ (which was, in fact, done above), but rather w.r.t.
the quantity $n_A$ which is connected to $\dot{z}^A$ by relations \eqref{ra22.1} and \eqref{ra23}.
Transforming the integration variables in the path integral (\ref{int}) from $\pi_A$ to $n_A$
{and using \eqref{ra22} and \eqref{ra23.1}}, we can rewrite {it in the form}
\begin{multline}\label{itog}
I=\int{Dz}{Dn}\left|\dfrac{\de \pi_A}{\de n_B}\right| \text{exp}\left\{{i}\int{d}y
\left(\dot{z}^A\left(\tr w_{AB}B^{BE}n_E-
\frac{1}{2}n_A\left(n_D B^{DE}n_E-\tr{w}_{DE} B^{DE}\right)\right)\right.\right.-\\
-\left.\left.\frac{1}{2}\sqrt{1-n_A\tr{w}^{AB}n_B}\left(n_D B^{DE} n_E-\tr{w}_{DE} B^{DE}\right)\right)\right\}.
\end{multline}

The Jacobian  $\left|\dfrac{\de \pi_A}{\de n_B}\right|$ which is present here is given by a quite simple expressions which is quadratic w.r.t. $n_A$:
\disn{ss1}{
	\frac{\dd \pi_A}{\dd n_B}=
	\ls \tr{w}_{AD}-n_A n_D\rs B^{DB}-\frac{1}{2} \de_A^B \ls n_D B^{DE} n_E - \tr{w}_{DE} B^{DE}\rs.
	\nom}

As usual, this Jacobian can be then rewritten as an additional ghost contribution to the Lagrangian with corresponding integration w.r.t. ghost fields.
Note that the path integral, when written in the form (\ref{itog}),  does not depend on any implicitly defined function.

\section{Conclusions}
In this work we discuss the approach to gravity which generalizes GR and has some potential advantages over GR from the point of view of the quantization. Among these advantages there is the fact that that the embedding theory, when formulated as splitting gravity, resembles some field theory in the higher-dimensional Minkowski space. It solves the causality problem, since the metric $\eta_{ab}$ of the ambient space is not altered by quantization (in contrast with GR, where the only metric $g_{\mu\nu}$ becomes an operator). It also solves the problem of time, since the time-like direction $y^0$ of the ambient space turns out to be the natural choice.

We study the canonical formulation of the considered theory, since it is one of the steps towards canonical (i.e., Dirac) quantization of a theory, in which we impose the canonical commutation relations on the generalized coordinates and momenta. Furthermore, for such complicated theories as gravity, the canonical formulation plays an important role even in the obtaining of $S$-matrix through path integral quantization. The reason of its importance lies in the fact that it is the path integral w.r.t. {canonical variables} of type \eqref{int}  that is equivalent to canonical quantization \cite{popov} and therefore leads to the $S$-matrix that is unitary by default. The usual expression for path integral quantization can be obtained from it after the integration w.r.t. generalized momenta, which can be performed only in simple cases, of Hamiltonians which are quadratic w.r.t. momenta. If this is not the case, it is more correct to use the path integral w.r.t. canonical variables.

It should be noted that the states, between which the amplitude \eqref{int} is defined, are written in terms of independent variable $z^A(y^a)$ of the splitting theory and not in terms of any functions of the coordinates $x^\mu$ on the surfaces (we remind that in the reformulation of gravity as splitting theory such coordinates are unnecessary).
However, one can establish a connection between these states and the states that defined in terms of usual variables.

The in and out states in 
\eqref{int} are some functions of $z'^A(y^a)$, $z''^A(y^a)$, which play the role of boundary conditions
\disn{za1}{
	z^A(y^a)\str{y^0\to-\infty} z'^A(y^a),\qquad
	z^A(y^a)\str{y^0\to\infty} z''^A(y^a)
	\nom}
for the function $z^A(y^a)$, with respect to which the path integration is performed
Each function $z'^A(y^a)$, $z''^A(y^a)$ defines a splitting of the ambient space
(in the regions $y^0\to-\infty$ and $y^0\to\infty$ correspondingly) into a system of four-dimensional surfaces, which have certain invariant geometric characteristics (curvature etc.). Therefore \eqref{int} can be interpreted as an amplitude of probability that at the given geometric characteristics in the past (corresponding to $z'^A(y^a)$) our spacetime will have certain geometric characteristics (corresponding to $z''^A(y^a)$) in the future.

If one requires that the functions $z'^A(y^a)$, $z''^A(y^a)$ in some sense correspond to a free (i.e., non-interacting) theory, as it usually supposed for the asymptotic states,
then the surfaces defined by them would have the metric corresponding to weak gravitational waves, so  \eqref{int} could be interpreted as the scattering amplitude of gravitons. 
Needless to say, many questions would arise on the way of realizing such ideas of interpretation of splitting gravity, which have to be understood and~studied.

Despite the fact that the path integral
(\ref{itog}) does not contain any implicitly defined function anymore, the resulting expression is, nevertheless, quite complicated.
{Some additional assumptions might be considered in a detailed analysis of this path integral, which lies beyond the scope of the present paper.}
For example, one can consider the Friedmann symmetry, when the ambient space can be chosen as five-dimensional (i.e., $N=5$), and $z^A(y^a)$ turns out to be scalar $z(y^a)$ \cite{statja43}. Another possibility is a non-relativistic approximation, when $n^A \ll 1$, and one can separate a Gaussian part of this integral. Finally, it is possible to consider some low-dimensional systems instead of four-dimensional surfaces, e.g., a two-dimensional string. The study of the description of a string can serve as a good test of this~formulation.

{\bf{Acknowledgments}}. {The authors are grateful to D.A.~Grad for the idea of using the action \eqref{ss4}. The work is supported by RFBR Grant No.~20-01-00081.}



\end{document}